\documentclass[a4paper,12pt]{article}
\usepackage{epsfig}
\usepackage[dvips,usenames]{color}
\usepackage{graphicx}

\newlength{\dinwidth}
\newlength{\dinmargin}
\def\pslash{\rlap{\hspace{0.02cm}/}{p}}
\def\eslash{\rlap{\hspace{0.02cm}/}{\epsilon}}

\begin{document}
\title{The production of the new gauge boson $B_{H}$ via
$e^{-}\gamma$ collision in the littlest Higgs model}
\bigskip
\author{Xuelei Wang\footnote{Email Address: wangxuelei@sina.com}, Zhenlan Jin, Qingguo Zeng\\
 {\small College of Physics and Information Engineering,}\\
 \small{Henan Normal
University, Xinxiang, Henan, 453007. P.R.China}
\thanks{This work is supported by the National Natural Science
Foundation of China(Grant No.10375017 and No.10575029).}
\\}
 \maketitle
\begin{abstract}

\indent The new lightest gauge boson $B_H$ with mass of a few
hundred GeV is predicted in the littlest Higgs model. $B_H$ should
be accessible in the planed ILC and the observation of such
particle can strongly support the littlest Higgs model. The
realization of $\gamma\gamma$ and $e\gamma$ collision will open a
wider window to probe $B_H$. In this paper, we study the new gauge
boson $B_{H}$ production processes $e^{-}\gamma\rightarrow
e^{-}\gamma B_{H}$ and $e^{-}\gamma\rightarrow e^{-}Z B_{H}$ at
the ILC. Our results show that the production cross section of the
process $e^{-}\gamma\rightarrow e^{-}Z B_{H}$ is less than one fb
in the most parameter spaces while the production cross section of
the process $e^{-}\gamma\rightarrow e^{-}\gamma B_{H}$ can reach
the level of tens fb and even hundreds of fb in the sizable
parameter spaces allowed by the electroweak precision data. With
the high luminosity, the sufficient typical signals could be
produced, specially via $e^{-}\gamma\rightarrow e^{-}\gamma
B_{H}$. Because the final electron and photon beams can be easily
identified and the signal can be easily distinguished from the
background produced by $Z$ and $H$ decaying, $B_H$ should be
detectable via $e\gamma$ collision at the ILC.  Therefore, the
processes $e^{-}\gamma\rightarrow e^{-}\gamma B_{H}$ and
$e^{-}\gamma\rightarrow e^{-}Z B_{H}$ provide a useful way to
detect $B_{H}$ and test the littlest Higgs model.
\end{abstract}
PACS number(s): 12.60.Cn, 14.80.Cp,13.66.Hk,11.30.Na
\newpage
\noindent{\bf I. Introduction}~~\\
\hspace{0.6cm}

The Standard Model (SM) of particle physics is a remarkably
successful theory. It provides a complete description of physics
at currently accessible energy, and its predictions have been
confirmed to high accuracy by the high energy experiments.
However, the mechanism of electroweak symmetry breaking(EWSB)
remains unknown. Furthermore, in the SM, the Higgs mass receives
quadratically divergent quantum corrections which have to be
cancelled by some new physics(NP) to avoid fine-tuning. The SM
Higgs sector is therefore an effective theory below some cut-off
scale $\Lambda$. To avoid fine-tuning of Higgs mass, one would
require the NP scale $\Lambda$ to be $\sim$ TeV. Various NP models
were proposed at the TeV scale, which can cancel the quadratic
divergences of the SM Higgs. Recently, a new theory, dubbed the
little Higgs theory\cite{LH}, has drawn a lot of interests as a
new candidate for solution of the problems mentioned above.

 \indent
 So far, a number of specific little Higgs models \cite{higgs-1,
 higgs-2,higgs-3,littlest} have been
 proposed. The generic structure of these models is that a global symmetry is
 broken at the scale f which is around a TeV. At the scale f there are new
 gauge bosons, scalars, fermions responsible for canceling the one
 loop quadratic divergences to the Higgs mass from the SM particles. The new
 particles predicted by the little Higgs models will emerge
 characteristic signatures at the present or future high energy
 collider experiments\cite{phenomena1}. Furthermore, among the various little
 Higgs models, the most economical and phenomenologically viable
 model is the littlest Higgs model\cite{littlest} which realizes
 the little Higgs idea and has all essential features of the little
 Higgs models. Such model consists of a nonlinear $\sigma$ model with
 a global SU(5) symmetry which is broken down to SO(5) by a vacuum
 expectation value (vev) of order f $\sim \Lambda_{s}/4\pi \sim$
 TeV. At the same time, the gauge subgroup $[SU(2)\times U(1)]^{2}$ is
 broken to its diagonal subgroup $SU(2)\times U(1)$, identified as the
 SM electroweak gauge group. This breaking scenario gives rise to
 four massive gauge bosons $(B_{H},Z_{H},W^{\pm}_{H})$. Thus, studying the possible signatures of the
 new gauge bosons and their contribution to some processes at
 high-energy colliders is a good method to test the littlest Higgs
 model and furthermore to probe the EWSB mechanism. However, in the
 littlest Higgs model, the masses of these new heavy gauge bosons
 are in the range of a few TeV, except for the mass of $B_{H}$ in the range of
 hundreds GeV. In fact, the gauge boson $B_{H}$ is the
 lightest new particle in the littlest Higgs model so that it should
 be the first signal at future experiments and would play the most important role
 in probing the littlest Higgs model.

\indent It is well known that the LHC  can directly probe the
possible NP beyond the SM up to a few TeV. If the new particles or
interactions will be directly discovered at the future hadron
collider experiments, the linear $e^{+}e^{-}$ collider(LC) will
then play a crucial role in the detailed and thorough study of
these new phenomena and in the reconstruction of the underlying
fundamental theories. In addition $e^{+}e^{-}$ physics, the future
LC provides a unique opportunity to study $\gamma\gamma$ and
$e\gamma$ interactions at high energy and luminosity comparable to
those in $e^{+}e^{-}$ collisions\cite{photon-collider}. High
energy photons for $\gamma\gamma$, $e\gamma$ collisions(Photon
Collider) can be obtained using Compton backscattering of laster
light off the high energy electrons. Such possibilities will be
realized at the International Linear Collider(ILC), with the
center of mass(c.m.) energy $\sqrt{s}$=300 GeV-1.5 TeV and the
yearly luminosity 500 $fb^{-1}$\cite{ILC}. The physics potential
of the Photon Collider is very rich and complements the physics
program of the $e^{+}e^{-}$ mode. The Photon Collider will
considerably contribute to the detailed understanding of new
phenomena, and in some scenarios it is the best instrument for the
discovery of elements of NP. In particular, the $e^{-}\gamma$
collision can produce particles which are kinematically not
accessible at $e^{+}e^{-}$ collisions at the same
collider\cite{photon-collider, collider}. Moreover, the high
energy photon polarization can vary relatively easily, which is
advantageous for experiments. All the virtues of the Photon
Collider will provide us with a good chance to pursue NP
particles, specially the lightest new gauge boson $B_H$ which
should be kinematically accessible at the planned ILC. Some $B_H$
production processes have been studied at the Photon
Collider\cite{chen, yue}.  In the reference\cite{chen}, we have
studied a process of the $B_{H}$ production associated with W
boson pair via photon-photon collision, i.e.,
$\gamma\gamma\rightarrow W^+W^-B_H$. Such process would offer a
good chance to probe the $B_H$ signal and to study the triple and
quartic gauge couplings involving $B_H$ and the SM gauge bosons
which shed important light on the symmetry breaking features of
the littlest Higgs model. Via $e\gamma$ collision, a single heavy
new gauge boson in the littlest Higgs model can also be produced
at the TeV energy colliders\cite{yue}. We find that there exist
another important $B_H$ production processes via $e\gamma$
collision at the ILC, i.e., $e^{-}\gamma\rightarrow e^{-}\gamma
B_{H}$ and $e^{-}\gamma\rightarrow e^{-}Z B_{H}$. In this paper,
we study the possibility of detecting the $B_H$ via these
processes and complement the probe of the littlest Higgs model.

\indent  This paper is organized as follows. In Sec. II, we
briefly review the littlest Higgs model. The Sec. III presents the
calculations of the production cross sections of the processes.
The numerical results and conclusions will be shown in Sec. IV.

\noindent{\bf II. The littlest Higgs model}~~\\
\hspace{0.6cm}

In this section we describe the main idea of littlest Higgs
model\cite{littlest} and the detailed review of this model can be
found in refrence\cite {phenomena1}. Furthermore, the
phenomenology of this model was also discussed in great detail in
precision tests and low energy
measurements\cite{han,heather,measurements,phenomena2}.

 The littlest Higgs model embeds the electroweak sector of the SM in
an SU(5)/SO(5) non-linear sigma model. The breaking of the global
$SU(5)$ symmetry to an $SO(5)$ subgroup at the scale
$\Lambda_{s}\sim 4\pi f$ by a vev of order f, results in 14
Goldstone bosons, which are denoted by $\Pi^{a}(x)$. We can
conveniently parameterize the Goldstone bosons by the non-linear
sigma model field
\begin{eqnarray}
\Sigma(x)=e^{i\Pi/f}\Sigma_0e^{i\Pi^{T}/f}=e^{2i\Pi/f}\Sigma_0,
\end{eqnarray}
where $f$ is decay constant, and
$\Pi(x)=\sum^{14}_{a=1}\Pi^{a}(x)X^{a}$. The sum runs over the 14
broken SU(5) generators $X^{a}$ and here we have used the relation
 $X^{a}\Sigma_{0}=\Sigma_{0}X^{aT}$, obeyed by the broken
 generators, in the last step.

 The leading order dimension-two term in the non-linear sigma model
 can be written for the scalar sector as
\begin{eqnarray}
\mathcal {L}_{\Sigma}=\frac{f^2}{8}Tr|D_{\mu}\Sigma|^2,
\end{eqnarray}
where the covariant derivative is
\begin{eqnarray}
 D_{\mu}\Sigma=\partial_{\mu}\Sigma-i\sum^{2}_{j=1}[g_j(W_{\mu j}\Sigma+\Sigma W^T_{\mu j})+
 g'_j(B_{\mu j}\Sigma+\Sigma B^T_{\mu j})].
\end{eqnarray}
$W_{\mu j},~B_{\mu j}$ are the $SU(2)_{j}$ and $U(1)_{j}$ gauge
fields, respectively and $g_{j},~ g'_{j}$ are the corresponding
coupling constants.

Furthermore, the vev breaks the gauge subgroup $[SU(2)\times
U(1)]^{2}$ of the SU(5) down to the diagonal group $SU(2)\times
U(1)$, identified as the SM electroweak group. So four of fourteen
Goldstone bosons are eaten to give mass to four particular linear
combinations of the gauge fields
\begin{eqnarray}
W=sW_1+cW_2, ~~~~W'=-cW_1+sW_2,\\ \nonumber
B=s'B_1+c'B_2,~~~~B'=-c'B_1+s'B_2,
\end{eqnarray}
with the mixing angle
\begin{eqnarray}
s\equiv sin\psi=\frac{g_2}{\sqrt{g^2_1+g^2_2}},~~~~c\equiv
cos\psi=\frac{g_1}{\sqrt{g^2_1+g^2_2}},\\ \nonumber
 s'\equiv sin\psi^{\prime}=\frac{g'_2}{\sqrt{g'^2_1+g'^2_2}}
,~~~~c'\equiv cos\psi^{\prime}=\frac{g'_1}{\sqrt{g'^2_1+g'^2_2}}.
\end{eqnarray}
These couplings can be related to the SM couplings ($g,g'$) by
\begin{eqnarray}
\frac{1}{g^2}=\frac{1}{g^2_{1}}+\frac{1}{g^2_{2}},~~~~
\frac{1}{g'^2}=\frac{1}{g'^2_{1}}+\frac{1}{g'^2_{2}}.
\end{eqnarray}
At the scale $f$, the SM gauge bosons $W$ and $B$ remain massless
while the heavy gauge bosons acquire masses of order $f$
\begin{eqnarray}
m_{W'}=\frac{g}{2s c}f, ~~~~  m_{B'}=\frac{g'}{2\sqrt{5}s'c'}f
\end{eqnarray}
The presence of $\sqrt{5}$ in the denominator of $m_{B'}$ leads to a
relatively light new neutral gauge boson.

The Higgs boson at tree level remains massless as a Goldtone
boson, but its mass is radiatively generated because any
nonlinearly realized symmetry is broken by the gauge, Yukawa, and
self-interactions of the Higgs field. The little Higgs model
introduces a collective symmetry breaking: Only when multiple
gauge symmetries are broken is the Higgs mass radiatively
generated; the loop corrections to the Higgs boson mass occur at
least at the two-loop level. The one-loop quadratic divergence
induced by the SM particles is canceled by that induced by the new
particles due to the exactly opposite couplings. For example, at
leading order in $1/f$, the couplings of H field to the gauge
bosons following from Eq.(2) are given as
\begin{eqnarray}
\mathcal {L}&=&\frac{1}{4}H(g_{1}g_{2}W_{1}^{\mu a}W^{a}_{2
\mu}+g'_{1}g'_{2}B^\mu_{1}B_{2\mu})H^{+}+...\\ \nonumber
&=&\frac{1}{4}H[g^{2}(W^{a}_{\mu}W^{\mu a}-W'^{a}_{\mu}W'^{\mu
a})-g'^{2}(B_{\mu}B^{\mu}-B'_{\mu}B'^{\mu})]H^{+}+...~.
\end{eqnarray}
It is to be compared with SUSY models where the cancellation occurs
due to the different spin statistics between the SM particle and its
superparter.

In order to cancel the severe quadratic divergence from the top
quark loop, another top-quark-like fermion is also required. In
addition, this new fermion is naturally expected to be heavy with
mass of order f. As a minimal extension, a vectorlike fermion pair
$\tilde{t}$  and   $\tilde{t}'^{c}$ with the SM quantum numbers
 $(3,1)_{Y_{i}}$ and  $(\bar{3},1)_{-Y_{i}}$  are introduced. With
  $ \chi_{i}=(b_{3},t_{3},\tilde{t})$ and antisymmetric tensors $\epsilon_{ijk}$
and $\epsilon_{xy}$, the
 coupling of the SM top quark to the pseudo-Goldstone bosons and the
 heavy vector pair in the littlest Higgs model is chosen to be
\begin{eqnarray}
\mathcal
{L}&=&\frac{1}{2}\lambda_{1}f\epsilon_{ijk}\epsilon_{xy}\chi_{i}\Sigma_{jx}\Sigma_{ky}u'^{c}_{3}+
\lambda_{2}f\tilde{t}\tilde{t'^{c}}+h.c. \\ \nonumber
&=&-i\lambda_{1}(\sqrt{2}h^{0}t_{3}+if\tilde{t}-\frac{i}{f}h^{0}h^{0*}\tilde{t})u'^c_{3}+h.c.+...~.
\end{eqnarray}
 As it is shown in the above equation, the quadratic divergence from the top quark
 is canceled by that from the new heavy top-quark-like fermion. In addition, the
 cancellation is stable against radiative corrections.

 The EWSB is induced by the remaining Goldstone bosons H and $\phi$.
 Through radiative corrections, the gauge, the Yukawa, and
 self-interactions of the Higgs field generate a Higgs potential
 which triggers the EWSB. Now the SM W and Z bosons acquire masses
 of order v, and small (of order $v^2/f^2$) mixing between the heavy
 gauge bosons and the SM gauge bosons W and Z occurs. The masses of
 the SM gauge bosons W and Z and their couplings to the SM particles
 are modified from those in the SM at the order of $v^2/f^2$.
  In the
 following, we denote the mass eigenstates of SM gauge fields by
 $W_{L}^{\pm}, Z_{L}$ and  $A_{L}$ and the new heavy gauge bosons by $W_{H}^{\pm}, Z_{H}$ and
$B_{H}$. The masses of the neutral gauge bosons are given to
$\mathcal
 {O}(\nu^{2}/f^{2})$  by \cite{han,Conley}
\begin{eqnarray}
M^{2}_{A_{L}}&=&0, \\ \nonumber
M^{2}_{B_{H}}&=&(M^{SM}_{Z})^{2}s^{2}_{W}\{\frac{f^{2}}{5s'^2c'^2v^2}-1
+\frac{v^2}{2f^2}[\frac{5(c'^2-s'^2)^2}{2s^2_W}-\chi_H\frac{g}{g'}\frac{c'^2s^2+c^2s'^2}{cc'ss'}]\},
\\ \nonumber
M^{2}_{Z_{L}}&=&(M^{SM}_{Z})^{2}\{1-\frac{v^{2}}{f^{2}}[\frac{1}{6}+\frac{1}{4}(c^{2}-s^{2})^{2}+
\frac{5}{4}(c'^{2}-s'^{2})^{2}]+8\frac{v'^2}{v^2}\},
\\ \nonumber
M^{2}_{Z_{H}}&=(&M^{SM}_{W})^{2}\{\frac{f^2}{s^2c^2v^2}-1+\frac{v^2}{2f^2}[\frac{(c^2-s^2)^2}{2c_W^2}
+\chi_H\frac{g'}{g}\frac{c'^2s^2+c^2s'^2}{cc'ss'}]\}.
\end{eqnarray}
Where
$\chi_{H}=\frac{5}{2}gg'\frac{scs'c'(c^{2}s'^{2}+s^{2}c'^{2})}{5g^{2}s'^{2}c'^{2}-g'^2s^{2}c^{2}}$,
$v$=246 GeV is the elecroweak scale, $v'$ is the vev of the scalar
$SU(2)_{L}$ triplet and $s_{W}(c_{W})$ represents the sine(cosine)
of the weak mixing angle.

The phenomenology of the littlest Higgs model at high energy
colliders depends on the following parameters:
\begin{eqnarray*}
f,~~~c,~~~c^{'},~~~x_{\lambda}.
\end{eqnarray*}
$x_{\lambda}=\lambda_{1}/\lambda_{2}$, and one of $\lambda_{1}$,
$\lambda_{2}$ can be replaced by the top-quark mass. Global fits
to the experimental data put rather severe constrains on
 the $f>4$ TeV at $95\%$ C.L. \cite{constrains}. However, their
 analyses are based on a simple assumption that the SM fermions are
 charged only under $U(1)_{1}$. If the SM fermions are charged under
 $U(1)_{1}\times U(1)_{2}$, the bounds become relaxed. The scale
 parameter $f= 1\sim 2$ TeV is allowed for the mixing parameters
 $c$ and $c'$ in the range of $0\sim0.5$ and
 $0.62\sim0.73$, respectively\cite{limit}.\\

 \noindent{\bf III. The cross sections for the
processes $e^{-}\gamma\rightarrow
e^{-}\gamma B_{H}, e^{-}Z B_{H}$}~~\\

 \indent Taking account of the gauge invariance of the
Yukawa couplings, one can write the couplings of the neutral gauge
bosons $\gamma,~Z$ and $B_{H}$ to the electron pair in the form of
$i\gamma^{\mu}(g_{V}+g_{A}\gamma^{5})$ \cite{han,Buras} with
\begin{eqnarray}
 g^{\gamma \bar{e}e}_V &=& -e ~,~~~~~~~~~~~~~~g^{\gamma
 \bar{e}e}_A=0,\\ \nonumber
g_{V}^{Zee}&=&-\frac{g}{2c_{W}}\{(-\frac{1}{2}+2s^{2}_{W})-\frac{v^2}{f^2}[-c_w\chi_Z^{W'}c/2s
+\frac{s_w\chi_Z^{B'}}{s'c'}(2y_e-\frac{9}{5}+\frac{3}{2}c'^2)]\},
 \\
\nonumber
g_{A}^{Zee}&=&-\frac{g}{2c_{W}}\{\frac{1}{2}-\frac{v^2}{f^2}[c_W\chi_Z^{W'}c/2s
+\frac{s_W\chi_Z^{B'}}{s'c'}(-\frac{1}{5}+\frac{1}{2}c'^2)]\},
 \\
\nonumber
g_{V}^{B_{H}ee}&=&\frac{g'}{2s'c'}(2y_e-\frac{9}{5}+\frac{3}{2}c'^{2}),
\\
\nonumber
g_{A}^{B_{H}ee}&=&\frac{g'}{2s'c'}(-\frac{1}{5}+\frac{1}{2}c'^{2}).
\end{eqnarray}
Where, $\chi_Z^{B'}=\frac{5}{2s_W}s'c'(c'^2-s'^2)$ and
$\chi_Z^{W'}=\frac{1}{2c_W}sc(c^2-s^2)$. The U(1) hypercharge of
electron, $y_e$, can be fixed by requiring that the $U(1)$ charge
assignments is anomaly free, i.e.,
   $y_e=\frac{3}{5}$. This is only one example among several alternatives for
the U(1) charge choice\cite{han}.

  With the above couplings, $B_H$ can be produced associated with a $\gamma$ or
   Z boson via $e^- \gamma$ collision. At the tree-level, the relevant Feynman
diagrams for the processes $e^{-}\gamma\rightarrow e^{-}\gamma
B_{H}, ~e^{-}Z B_{H}$ in the littlest model  are shown in
Figs.1(a-f).
\begin{figure}[h]
\begin{center}
\epsfig{file=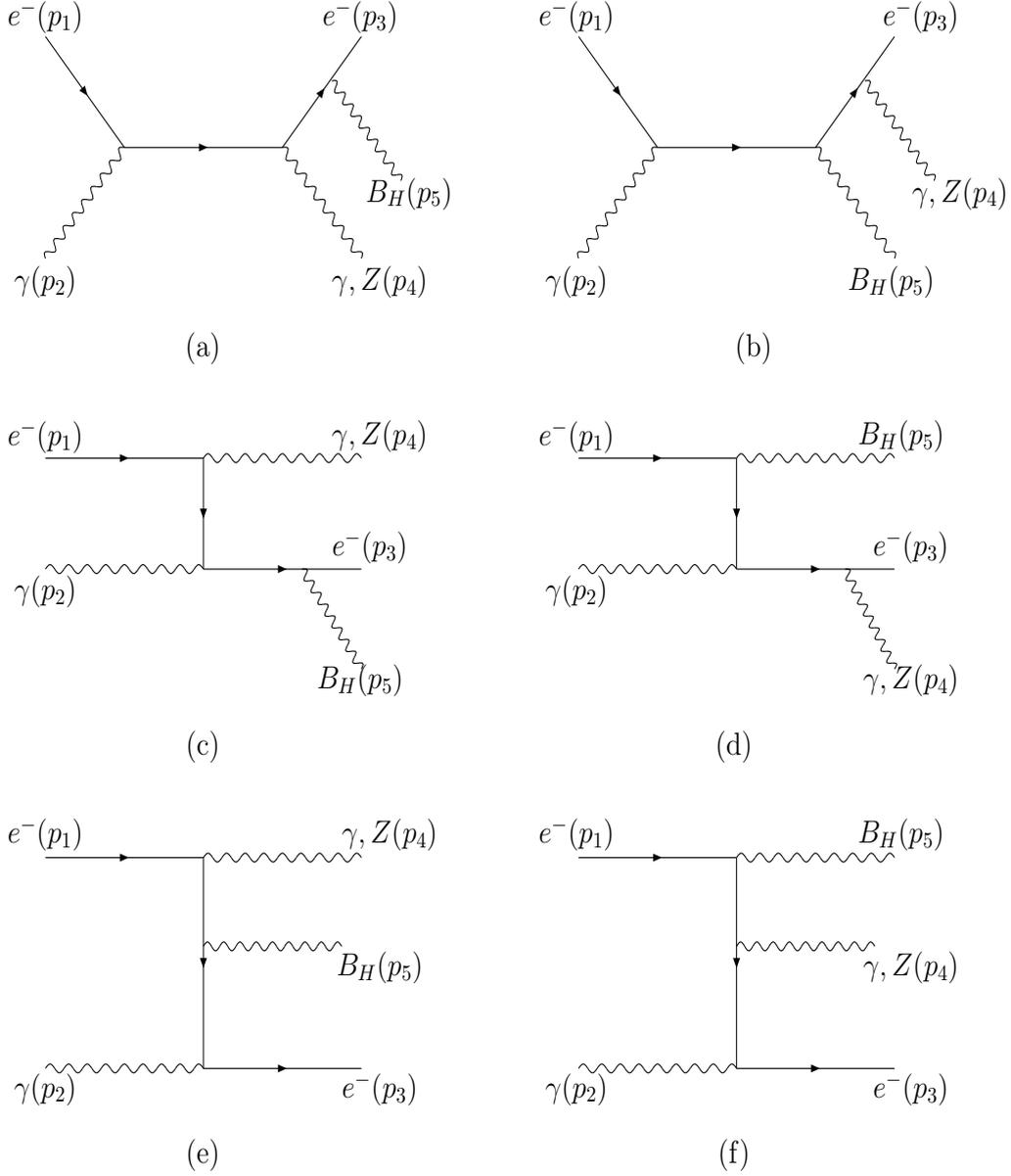,width=450pt,height=720pt} \vspace{-6.5cm}
\caption{\small The Feynman diagrams of the processes
$e^{-}\gamma\rightarrow e^{-}\gamma B_{H},~ e^{-}Z B_{H}$ in the
littlest Higgs model.}
\end{center}
\end{figure}
 In our calculation, we will neglect the electron mass and
define some notations as follows
 \begin{eqnarray}
 G(p)&=&\frac{1}{p^{2}},\\ \nonumber
  \Lambda^{V_i\bar{e}e}&=&g^{V_i\bar{e}e}_V+g^{V_i\bar{e}e}_A\gamma^{5},\\
  \nonumber
 \Lambda^{B_H\bar{e}e}&=&g^{B_H\bar{e}e}_V+g^{B_H\bar{e}e}_A\gamma^{5},
  \end{eqnarray}
 where $V_i$ presents the SM gauge bosons $\gamma,~Z$ and G(p) denotes the
 propagator of the electron.
 The production amplitudes of the processes can be
written as
\begin{eqnarray}
M_{V_i}=M^a_{V_i}+M^b_{V_i}+M^c_{V_i}+M^d_{V_i}+M^e_{V_i}+M^f_{V_i},
\end{eqnarray}
with
 \begin{eqnarray}
 M^a_{V_i}=G(p_1+p_2)G(p_3+p_5)\bar{u}_e(p_3)\eslash(p_5)\Lambda^{B_H\bar{e}e}(\pslash_3+\pslash_5)
 \eslash(p_4)\Lambda^{V_i\bar{e}e}(\pslash_1+\pslash_2)\eslash(p_2)\Lambda^{\gamma\bar{e}e}u_e(p_1),\nonumber
 \end{eqnarray}
\begin{eqnarray}
M^b_{V_i}=G(p_1+p_2)G(p_3+p_4)\bar{u}_e(p_3)\eslash(p_4)\Lambda^{V_i\bar{e}e}(\pslash_3+\pslash_4)
 \eslash(p_5)\Lambda^{B_H\bar{e}e}(\pslash_1+\pslash_2)\eslash(p_2)\Lambda^{\gamma\bar{e}e}u_e(p_1),\nonumber
 \end{eqnarray}
\begin{eqnarray}
M^c_{V_i}=G(p_1-p_4)G(p_3+p_5)\bar{u}_e(p_3)\eslash(p_5)\Lambda^{B_H\bar{e}e}(\pslash_3+\pslash_5)
 \eslash(p_2)\Lambda^{\gamma\bar{e}e}(\pslash_1-\pslash_4)\eslash(p_4)\Lambda^{V_i\bar{e}e}u_e(p_1),\nonumber
 \end{eqnarray}
 \begin{eqnarray}
M^d_{V_i}=G(p_1-p_5)G(p_3+p_4)\bar{u}_e(p_3)\eslash(p_4)\Lambda^{V_i\bar{e}e}(\pslash_3+\pslash_4)
 \eslash(p_2)\Lambda^{\gamma\bar{e}e}(\pslash_1-\pslash_5)\eslash(p_5)\Lambda^{B_H\bar{e}e}u_e(p_1),\nonumber
 \end{eqnarray}
 \begin{eqnarray}
M^e_{V_i}=G(p_3-p_2)G(p_1-p_4)\bar{u}_e(p_3)\eslash(p_2)\Lambda^{\gamma\bar{e}e}(\pslash_3-\pslash_2)
 \eslash(p_5)\Lambda^{B_H\bar{e}e}(\pslash_1-\pslash_4)\eslash(p_4)\Lambda^{V_i\bar{e}e}u_e(p_1),\nonumber
 \end{eqnarray}
 \begin{eqnarray}
M^f_{V_i}=G(p_3-p_2)G(p_1-p_5)\bar{u}_e(p_3)\eslash(p_2)\Lambda^{\gamma\bar{e}e}(\pslash_3-\pslash_2)
 \eslash(p_4)\Lambda^{V_i\bar{e}e}(\pslash_1-\pslash_5)\eslash(p_5)\Lambda^{B_H\bar{e}e}u_e(p_1).
 \end{eqnarray}

 \indent The hard photon beam of the $e\gamma$ collider can be obtained
from laser backscattering at the $e^+e^-$ linear collider. Let
$\hat{s}$ and $s$ be the center-of-mass(c.m.) energies of the
$e\gamma$ and $e^+e^-$ systems, respectively. Using the above
amplitudes, we can directly obtain the cross sections
$\hat{\sigma}(\hat{s})$ for the sub-process $e^-\gamma\rightarrow
e^-\gamma B_{H}$ and $e^-\gamma\rightarrow e^-Z B_{H}$, and the
total cross sections at the $e^+e^-$ linear collider can be
obtained by folding $\hat{\sigma}(\hat{s})$ with the photon
distribution function $f_{\gamma}(x)$ which is given in
Ref.\cite{distribution}
\begin{eqnarray}                                                   
\sigma_{tot}(s)=\int\limits_{M_{final}^2/s}^{x_{max}}dx\hat{\sigma}
(\hat{s})f_{\gamma}(x).
\end{eqnarray}
$M_{final}$ is the sum of the masses of the final state particles
and
\begin{eqnarray}                                                    
\displaystyle f_\gamma(x)=\frac{1}{D(\xi)}\left[1-x+\frac{1}{1-x}
-\frac{4x}{\xi(1-x)}+\frac{4x^2}{\xi^2(1-x)^2}\right],
\end{eqnarray}
with
\begin{eqnarray}                                                    
\displaystyle D(\xi)=\left(1-\frac{4}{\xi}-\frac{8}{\xi^2}\right)
\ln(1+\xi)+\frac{1}{2}+\frac{8}{\xi}-\frac{1}{2(1+\xi)^2}\,.
\end{eqnarray}
In above equations, $\xi=4E_e\omega_0/m_e^2$ in which $m_e$ and
$E_e$ stand, respectively, for the incident electron mass and
energy, $\omega_0$ stands for the laser photon energy, and
$x=\omega/E_e$ stands for the fraction of energy of the incident
electron carried by the back-scattered photon. $f_\gamma$ vanishes
for $x>x_{max} =\omega_{max}/E_e=\xi/(1+\xi)$. In order to avoid
the creation of $e^+e^-$ pairs by the interaction of the incident
and back-scattered photons, we require $\omega_0x_{max}\leq
m_e^2/E_e$ which implies that $\xi\leq 2+ 2\sqrt{2}\approx 4.8$.
For the choice of $\xi=4.8$, we obtain
\begin{eqnarray*}                                                     
x_{max}\approx 0.83,\hspace{1cm} D(\xi)\approx 1.8 \,.
\end{eqnarray*}
For simplicity, we have ignored the possible polarization for the
electron and photon beams.\\

\noindent{\bf IV. The numerical results and conclusions }~~\\

 \indent To obtain numerical results, we take
 $M_Z=91.187$ GeV, $v=246$ GeV,  $s^{2}_{W}=0.23$.
  The
electromagnetic fine structure constant $\alpha_e$ at certain
energy scale is calculated from the simple QED one-loop evolution
formula with the boundary value $\alpha_e=1/137.04$
\cite{Donoghue}. There are four free parameters in our numerical
estimations, i.e., $f,c,c',\sqrt{s}$. Here, we take the parameter
space ($f=1\sim2$ TeV, $c=0-0.5$, $c'=0.62-0.73$) which is
consistent with the electroweak precision data. The final
numerical results are summarized in Figs.2-3.
\begin{figure}[h]
\begin{center}
\scalebox{1.25}{\epsfig{file=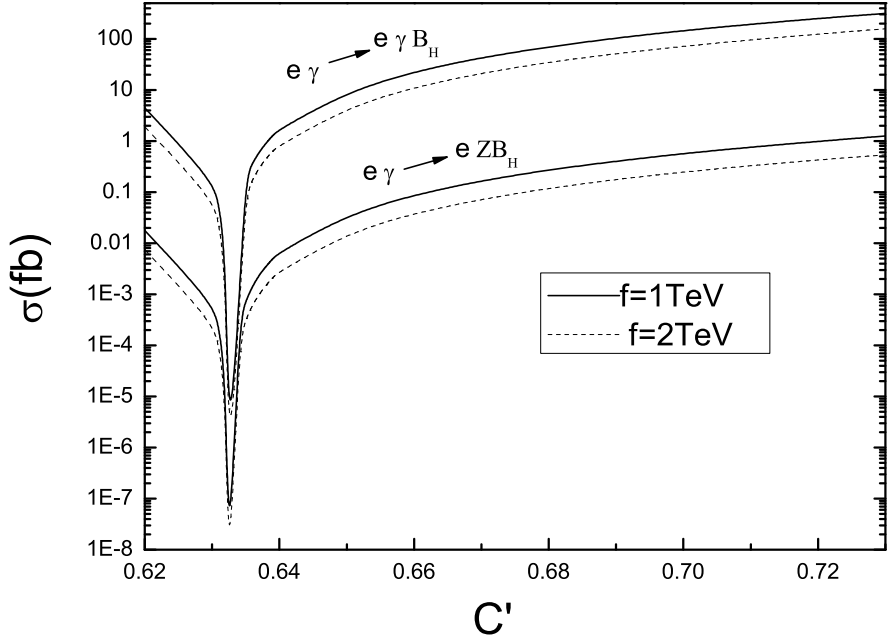}}\\
\end{center}
\caption{\small The production cross sections of the processes
$e^{-}\gamma\rightarrow e^{-}\gamma B_{H}$ (upper curves) and
$e^{-}\gamma\rightarrow e^{-}Z B_{H}$ (lower curves) as a function
of the mixing parameter $c'$  for $\sqrt{s}=800$ GeV and the scale
parameter f=1 TeV(solid line), and f=2 TeV(dashed line),
respectively.}
\end{figure}
\begin{figure}[h]
\begin{center}
\scalebox{1.25}{\epsfig{file=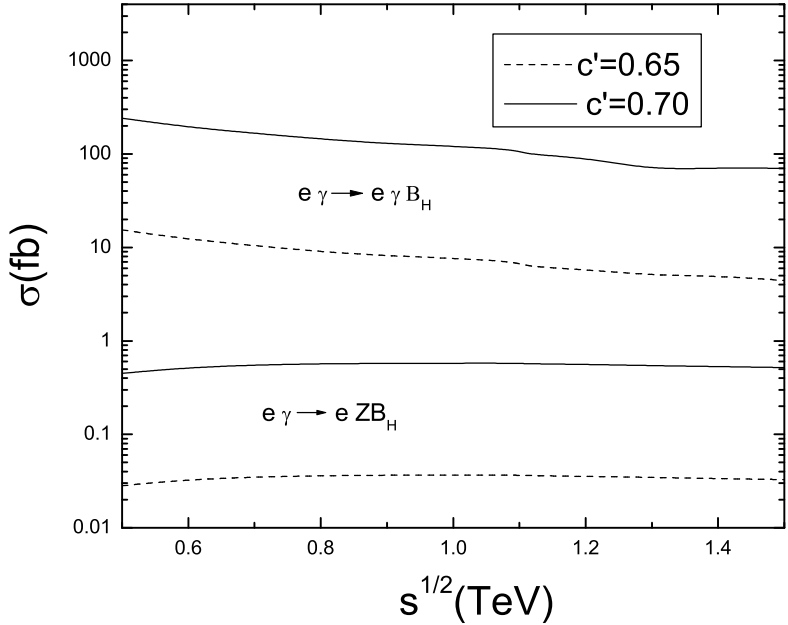}}\\
\end{center}
\caption{\small The production cross sections of the processes
$e^{-}\gamma\rightarrow e^{-}\gamma B_{H}$(upper curves) and
$e^{-}\gamma\rightarrow e^{-}Z B_{H}$(lower curves) as a function
of the c.m. energy $\sqrt{s}$  for f=1 TeV and the mixing
parameter c'=0.70(solid line), and c'=0.65(dashed line),
respectively.}
\end{figure}

 The production cross sections of the processes $e^{-}\gamma\rightarrow e^{-}\gamma B_{H}$ and
$e^{-}\gamma\rightarrow e^{-}Z B_{H}$ are not sensitive to c, and
we fix the value of c as 0.4 in our calculation. The cross
sections mainly depend on the mixing parameter $c'$. So in Fig.2
we plot the cross sections as a function of the parameter $c'$,
taking $\sqrt{s}=0.8$ TeV and f=1 TeV(2 TeV) as the examples. From
Fig.2, one can see that the cross sections drop sharply to zero
when $c'$ equals $\sqrt{2/5}$. This is because the coupling of the
gauge boson $ B_H$ to the electron pair become decoupled with
$c'=\sqrt{2/5}$. When $c'$ is over $\sqrt{2/5}$, the cross
sections increase with $c'$ and the cross section of the process
$e^{-}\gamma\rightarrow e^{-}\gamma B_{H}$ can reach the level of
tens fb even a hundred fb. But the cross section of the process
$e^{-}\gamma\rightarrow e^{-}Z B_{H}$ is much smaller than that of
$e^{-}\gamma\rightarrow e^{-}\gamma B_{H}$ and its maximum can
only reach the level of 1 fb. So the process
$e^{-}\gamma\rightarrow e^{-}\gamma B_{H}$  should have advantage
in probing $B_H$. On the other hand, comparing the results for f=1
TeV with those for f=2 TeV, we find that the cross sections
decrease slightly with f increasing. This is mainly because that
the mass of $B_H$ increases with f increasing which can depress
the phase space.

To show the influence of the c.m. energy $\sqrt{s}$ on the cross
sections, we plot the cross sections as a function of $\sqrt{s}$
with f=1 TeV and $c'=0.65, 0.70$ in Fig.3. Considering the c.m.
energy at the ILC and the kinetic limit, we present the numerical
results for energies ranging from
 0.5 to 1.5 TeV. The results show that the cross section of $e^- \gamma
 B_H$ production slightly decreases with $\sqrt{s}$¡¡
 and the cross section
 of $e^-Z B_H$ production is more insensitive to $\sqrt{s}$.

\indent  The yearly luminosity of the ILC can reach $500 fb^{-1}$.
So we can conclude that the sufficient typical $B_H$ events can be
produced via $e\gamma$ collision, specially via the process
$e^{-}\gamma\rightarrow e^{-}\gamma B_{H}$.¡¡
 But to detect $B_H$,
one also needs to study the decay modes of $B_H$ which has been
done in reference\cite{han}. The decay width of a particle affects
how and to what extent it is experimentally detectable, since
production of particles with very large decay widths may be
difficult to distinguish from background processes. For $B_H$, the
parameter spaces where the large decay width would occur are
beyond current search limits in any case. So if $B_H$ would be
produced it can be detected via the measurement of the peak in the
invariant mass distribution of its decaying particles.¡¡
 On the
other hand, for the process $e^{-}\gamma\rightarrow e^{-}\gamma
B_{H}$, the backgrounds are likely to be more larger in the
collision direction because it is difficult to distinguish the
final state $e^- \gamma$ from the injecting $e^- \gamma$. But the
backgrounds can be significantly depressed if the detection is
taken in the direction of deviating the collision.

 In the
following, we focus on discussing how to detect $B_H$ via its
decay modes. The main decay modes of $B_H$ are
$e^+e^-+\mu^+\mu^-+\tau^+\tau^-,~
d\bar{d}+s\bar{s},~u\bar{u}+c\bar{c},~ZH,~W^+W^-$. The decay
branching ratios of these modes have been studied in
reference\cite{han} which are strongly dependent on the $U(1)$
charge assignments of the SM fermions. In general, the heavy gauge
bosons are likely to be discovered via their decays modes to
leptons. For $B_H$, the most interesting decay modes should be
$e^+e^-,~\mu^+\mu^-$. This is because such leptons can be easily
identified and the number of $e^+e^-,~\mu^+\mu^-$ background
events with such a high invariant mass is very small. So, the
measurement of the peak in the invariant mass distribution of
$e^+e^-,~\mu^+\mu^-$ can provide a unique way to probe $B_H$. For
the signal $e^{-}\gamma e^+e^-(\mu^+\mu^-)$, the main SM
background arises from $e^{-}\gamma\rightarrow e^{-}\gamma Z$ with
$Z\rightarrow e^+e^-(\mu^+\mu^-)$. The cross section is a few pb
with $\sqrt{s}$ =0.5 $\sim$ 1.5 TeV \cite{erz}. For the signal $
e^{-}Ze^+e^-(\mu^+\mu^-)$, the most serious SM backgrounds come
from the processes $e^-\gamma\rightarrow e^-Z Z$,
$e^-\gamma\rightarrow e^-Z H$ and their cross sections can reach
about 10 fb, a few fb, respectively, in the energy range
$\sqrt{s}=0.5\sim 1.5$ TeV\cite{ezz}. But one can very easily
distinguish $B_H$ from $Z$ via their significantly different
$e^+e^-$($\mu\mu$) invariant mass distribution. The backgrounds of
$e^-\gamma\rightarrow e^-Z H$ with H decaying to lepton pair or
light quark pair are very small because the decay branching ratios
of these H decay modes are depressed by the small masses of
leptons and light quarks. On the other hand, we can also
distinguish $B_H$ from H via their different invariant mass
distribution of final particles because $B_H$ is much heavier than
H. As we know, in a narrow region around $c'=0.63$, the decay
branching ratios of $B_H\rightarrow l^+l^-$ approach zero due to
the decoupling of $B_H$ with lepton pair. In this case, the main
decay modes of $B_H$ are $B_H\rightarrow W^+W^-, ZH$. The decay
mode $Z\rightarrow W^+W^-$ is of course kinematically forbidden in
the SM but $H\rightarrow W^+W^-$ is the dominant decay mode with
Higgs mass above 135 GeV(one or both of W is off-shell for Higgs
mass below 2$M_W$). So the background for the signal $e^-ZW^+W^-$
might be serious and it is hard to detect the $B_H$ via
$e^-\gamma\rightarrow e^- ZB_H$ with $B_H\rightarrow W^+W^-$.
However, the process $e^-\gamma\rightarrow e^-\gamma B_H$ does not
suffer such large background problem which would be another
advantage of $e^-\gamma\rightarrow e^-\gamma B_H$. For
$B_H\rightarrow ZH$, the main final states are $l^+l^-b\bar{b}$.
In this case, two b-jets can be reconstructed to the Higgs mass
and a $l^+l^-$ can be reconstructed to the Z mass and the
background is very clean. Furthermore, the decay mode $ZH$
involves the off-diagonal coupling $HZB_H$ and the factor
$cot2\psi'$ in the coupling $HZB_H$ is a unique feature of the
littlest Higgs model. It should also be mentioned that
experimental precision measurement of such off-diagonal coupling
is more easier than that of diagonal coupling. So, the decay mode
$ZH$ would not only provide a better way to probe $B_H$ but also
provide a robust test of the littlest Higgs model.

 In conclusion, the realization of $e\gamma$ or $\gamma\gamma$ collision at
 the planed ILC with high energy and luminosity will provide more chances to
 probe the new particles predicted by the new physics beyond the SM.
 In this paper, we study the new gauge boson $B_H$ production
 processes via $e\gamma$ collision, i.e., $e^{-}\gamma\rightarrow
e^{-}\gamma B_{H},~e^{-}Z B_{H}$. The study shows that the cross
section of $e^{-}\gamma\rightarrow e^{-}Z B_{H}$ can reach the
level of less than one fb in most parameter spaces while the cross
section of the process $e^{-}\gamma\rightarrow e^{-}\gamma B_{H}$
can reach the level of tens fb and even hundreds of fb in most
parameter spaces allowed by the electroweak precision data. We can
predict that there are enough $B_H$ signals can be produced via
these processes, specially via $e^{-}\gamma\rightarrow e^{-}\gamma
B_{H}$ at the planned ILC. Because the new gauge boson $B_H$ can
be easily distinguished from the SM Z and H, the signal would be
typical and the background would be very clean. So, the processes
$e^{-}\gamma\rightarrow e^{-}\gamma B_{H},~e^{-}Z B_{H}$ will
provide a good chance to probe $B_H$ and test the littlest Higgs
model.

\end{document}